\documentclass[twocolumn]{aastex701}

\begin{document}

\title{Evolution of Massive Main-sequence Stars in Rapid Population Synthesis. I. Framework and Implementation}

\author[0009-0002-2040-0637]{Adam Br\v{c}ek}
\affiliation{School of Physics and Astronomy, Monash University, Clayton, VIC 3800, Australia}
\affiliation{OzGrav: The ARC Centre of Excellence for Gravitational Wave Discovery, Clayton, VIC 3800, Australia}
\email{adam.brcek@monash.edu}

\author[0000-0002-8032-8174]{Ryosuke Hirai}
\affiliation{Astrophysical Big Bang Laboratory (ABBL), Pioneering Research Institute (PRI), RIKEN, Wako, Saitama 351-0198, Japan}
\affiliation{School of Physics and Astronomy, Monash University, Clayton, VIC 3800, Australia}
\affiliation{OzGrav: The ARC Centre of Excellence for Gravitational Wave Discovery, Clayton, VIC 3800, Australia}
\email{ryosuke.hirai@monash.edu}

\author[0000-0002-6134-8946]{Ilya Mandel}
\affiliation{School of Physics and Astronomy, Monash University, Clayton, VIC 3800, Australia}
\affiliation{OzGrav: The ARC Centre of Excellence for Gravitational Wave Discovery, Clayton, VIC 3800, Australia}
\email{ilya.mandel@monash.edu}

\author{Harmony Lower}
\affiliation{School of Physics and Astronomy, Monash University, Clayton, VIC 3800, Australia}
\email{hlow0015@student.monash.edu}

\correspondingauthor{Adam Br\v{c}ek}
\email{adam.brcek@monash.edu}

\begin{abstract}

Stars spend most of their lifetime on the main sequence (MS), where hydrogen burning establishes the internal chemical structure that governs the subsequent evolution. In massive stars, mass loss through winds and binary interactions can significantly modify this structure during the MS. We present a new MS evolution framework suitable for rapid binary population synthesis, implemented in the COMPAS code. Building on the semi-analytical model of Shikauchi et al. (2025), our framework captures the evolution of the convective core on the MS under arbitrary mass-loss or mass-gain histories, including a treatment for stellar rejuvenation and MS mergers. This new framework yields more massive helium cores at terminal-age MS, more compact radii in stripped MS stars, and systematically higher black hole masses than commonly used prescriptions. By providing a more realistic treatment of MS evolution, this framework improves the physical consistency of massive stars and binary evolution in rapid population synthesis.

\end{abstract}



\section{Introduction} \label{sec:intro}

Stars spend most of their lifetimes on the main sequence (MS), the evolutionary stage characterized by core hydrogen burning. Mass loss or accretion during the MS can have major consequences for the star's subsequent evolution. In single stars, mass loss is driven by stellar winds. The most massive stars have particularly strong winds, which can strip away a significant fraction of their mass while still on the MS \citep[e.g.][]{Vink2022}. In binary systems, stars may exchange mass through Roche-lobe overflow. When this occurs while the donor is on the MS, the interaction is classified as case A mass transfer. Binary interactions and especially case A mass transfer are critical to account for in massive stars, as O-type stars (with masses $\gtrsim 15\;M_\odot$) are preferentially born in binaries or higher multiples, and approximately 25\% are expected to interact while on the MS \citep{Sana2012, Moe2017}.

Many rapid binary population synthesis codes simplify stellar evolution by using analytical fits to global quantities like surface luminosity and radius as a function of mass and age, based on a set of detailed single star evolution models. When the star loses mass, it simply switches to different stellar mass evolution track, erasing any evolutionary history. In such models, the helium core mass at terminal-age main sequence (TAMS; refers to the time of core hydrogen exhaustion) is determined solely by the total stellar mass. Consequently, two stars with the same mass at TAMS are assigned identical helium core masses, regardless of their mass-loss histories. This becomes problematic if one of the stars was initially more massive and lost a substantial amount of mass during the MS, particularly if the stripping occurs late in the MS when the star already has a significantly He-enriched core. In such cases, the helium core mass at TAMS becomes significantly underestimated.

The internal chemical structure established during the MS largely determines the star's subsequent evolution. This chemical distribution is set by the decay and growth of the mixed core, the central region in massive stars where efficient mixing occurs, which includes the convective core and the adjacent overshoot region. The mass of the mixed core at TAMS, determines the hydrogen-free core mass that later governs the outcome of core collapse. A semi-analytical model describing the time evolution of the mixed core mass under arbitrary mass-loss histories was recently developed by \citet{Shikauchi2024}. Throughout this work, we refer to the mixed core as the convective core.

In this work, we present a new MS evolution framework for massive stars that captures the evolution of the convective core. We extend the model for convective core mass evolution from \citet{Shikauchi2024} to account for mass gain by introducing a treatment for the rejuvenation of MS accretors. We describe the implementation of the new framework into the rapid binary population synthesis code \textsc{Compas}. We then explore the direct consequences of this new MS evolution framework, compare it with existing MS core treatments, and highlight its implications for massive single and binary star evolution.

\section{Method and theoretical framework}

\subsection{\textsc{Compas}} \label{sec:COMPAS}

\textsc{\textsc{Compas}}\footnote{Publicly available on \url{https://github.com/TeamCOMPAS/COMPAS}} (Compact Object Mergers: Population Astrophysics and Statistics, \citealt{Stevenson2017, Vigna-Gomez2018, Riley2021, Mandel2025}) is a rapid binary population synthesis code that models the evolution of binary stars using fits to more detailed models and their interpolations, allowing it to produce large binary populations for a low computational cost. The single star evolution is based on the analytic formulae of \citet{Hurley2000}, which are fits to the stellar models of \citet{Pols1998}. In this work, we use \textsc{Compas} v03.27.02 with default settings unless specified otherwise, including the default wind mass-loss prescription from \citet{Merritt2025} and stellar tides disabled. In the default mass-loss scheme of \citet{Merritt2025}, the \citet{Vink2021} prescription is applied to MS stars, \citet{Sabhahit2023} to very massive stars ($M > 100,M_\odot$), \citet{Decin2024} to red supergiants, and the prescriptions of \citet{Sander2023} and \citet{Vink2017} to Wolf--Rayet stars. We introduce a new \textsc{Compas} program option \texttt{--main-sequence-core-mass-prescription}, with modes \texttt{HURLEY}, \texttt{MANDEL}, and \texttt{BRCEK}.

The \texttt{HURLEY} option uses the original formalism from \citet{Hurley2000}, in which the internal structure of MS stars is not modeled. A mass-losing MS star moves onto a stellar track corresponding to its new mass, and MS stars are characterized only by their age and total mass. With this approach, MS stars do not retain any memory of mass loss, and the helium core mass is determined only from the total mass at TAMS.

Erased mass-loss history with the \texttt{HURLEY} MS evolution formalism can lead to significant underestimation of the helium core mass at TAMS after MS mass loss. This is partially addressed with the \texttt{MANDEL} option, which is a qualitative workaround to help MS stars retain a higher helium core mass at TAMS after case A mass transfer. With this method, the minimum core mass of the donor is set to the helium core mass of a Hertzsprung gap (HG) star with the same mass as the donor prior to mass transfer, scaled by the star's fractional MS age \citep{Romero-Shaw2023}. While this approach provides a systematically higher estimate for helium core masses at TAMS, it does not describe the full convective core mass evolution, does not take wind mass loss into account, and does not apply to MS accretors.

Throughout this work, we refer to our new framework that includes modeling of the convective core on the MS, rejuvenation, and updated stellar tracks as the \texttt{BRCEK} MS evolution framework. The details of the implementation are described in Section \ref{sec:implementation}. 

\subsection{Convective core mass evolution} \label{sec:CC_mass_evolution}
\citet{Shikauchi2024} introduce a semi-analytical model that describes the convective core mass evolution of massive stars that experience mass loss on the MS. Their model is based on 1D stellar evolution models from \textsc{Mesa} \citep{Paxton2011,Paxton2013,Paxton2015,Paxton2018,Paxton2019}. They identify several universal relations, which are used to construct the analytical framework consisting of solving two coupled differential equations describing the evolution of central helium fraction and the convective core mass for an arbitrary mass-loss history. The models by \citet{Shikauchi2024} adopt a treatment of convective boundary mixing and overshooting that differs from the approach used in the stellar models of \citet{Pols1998}, which underlie the fitting formulae of \citet{Hurley2000}. The \citet{Shikauchi2024} models use a mixing-length parameter $\alpha_\mathrm{mlt} = 2$ and apply step overshooting with an overshoot parameter $\alpha_\mathrm{ov}=0.2$, extending the mixing region by a fixed fraction of the pressure scale height. In contrast, \citet{Pols1998} do not parametrize the overshooting length in units of the pressure scale height, but use an approach that is based on the stability criterion itself. These differences in the treatment of convective boundary mixing affect stellar radii, luminosities, and MS lifetimes, and therefore lead to differences between the resulting stellar tracks.

The time evolution of the central helium abundance can be expressed as
\begin{equation} \label{eq:change_in_Yc_differential}
    \frac{\mathrm{d} Y_c}{\mathrm{d} t} = \frac{L}{Q_\mathrm{CNO} M_c},
\end{equation}
where $L$ and $M_c$ are the luminosity and convective core mass, and $Q_{\mathrm{CNO}}=6.019 \times 10^{18}$ erg $\mathrm{g}^{-1}$ is the energy released per unit mass by hydrogen fusion via the CNO cycle.

\citet{Shikauchi2024} model the time evolution of the convective core during the MS using
\begin{equation} \label{eq:core_change}
\frac{\mathrm{d}\:\mathrm{ln}\: M_c}{\mathrm{d} t} = -\frac{\alpha}{1-\alpha Y_c} \: \frac{\mathrm{d} Y_c}{\mathrm{d} t} + \beta \: \delta \frac{\dot{M}}{M},
\end{equation}
where $\dot{M}$ is the mass-loss rate, and $\alpha(M_c)$, $\beta(M)$ and $\delta(M_c,Y_c)$ are functions defined by \citet{Shikauchi2024}. The first term on the right-hand side of Eq.~(\ref{eq:core_change}) describes the natural shrinkage of the convective core mass as hydrogen is converted to helium on the MS. This decrease in convective core mass is accelerated by any additional mass loss that the star experiences, which is described by the second term in Eq.~(\ref{eq:core_change}).
The function $\delta$ describes the convective core's ``inertia", accounting for the fact that stars towards the end of MS are less willing to change their convective core.

The fitting functions $\alpha(M_c)$, $\beta(M)$, and $\delta(M_c,Y_c)$ contain coefficients that are listed for three metallicities: $Z=0.02$, $0.007$, and $0.002$. Since we want to use the new framework across a wide metallicity range, we linearly interpolate all coefficients in $\log Z$ between these calibration points and use constant extrapolation outside the range. This choice represents the simplest interpolation method that does not introduce additional assumptions about the functional dependence of coefficients on metallicity. Constant extrapolation beyond the calibrated metallicity range prevents the coefficients from being extrapolated to unphysical values. In this work, we adopt $Z_\odot=0.0142$ \citep{Asplund2009}.

\begin{figure}[b!]
\plotone{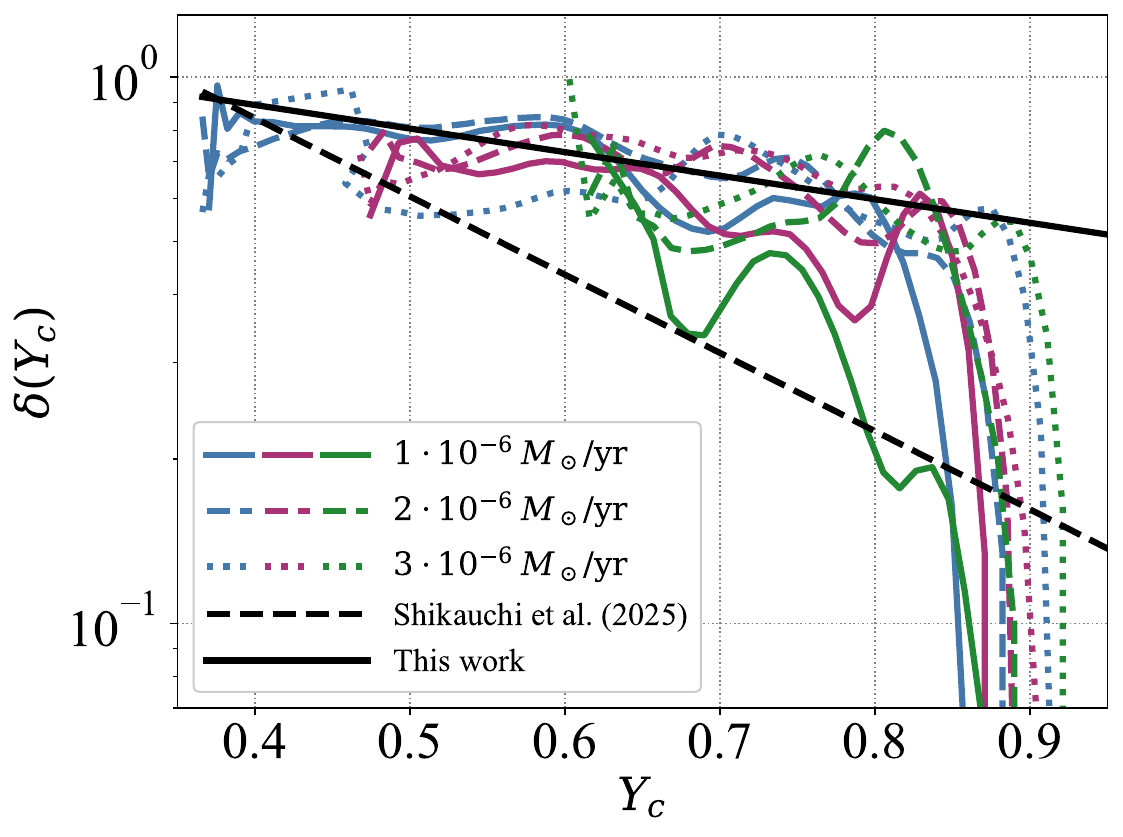}
\caption{Computed values of $\delta$ as a function of the central helium fraction $Y_c$ for a mass-gaining 40 $M_\odot$ MS star, based on our \textsc{Mesa} simulations. Three different mass gain onset times are shown: 0.88 Myr (blue), 1.76 Myr (purple), and 2.7 Myr (green). We apply 5-mean smoothing to the data. For each onset time, three mass gain rates are considered, as indicated in the legend.
\label{fig:new_delta_prescription}}
\end{figure}

The convective core mass framework of \citet{Shikauchi2024} accounts only for mass loss ($\dot M<0$) during the MS. The response of the convective core to mass gain is qualitatively different to mass loss, and there is no physical reason why the function $\delta$ should be the same in both regimes. We therefore use a set of \textsc{Mesa} simulations of accreting MS stars with initial masses of 15 $M_\odot$ and 40 $M_\odot$, following the approach of \citet{Shikauchi2024}, with inlists and subroutines publicly available \citep{shikauchi_2024_12662374}. We assume idealized mass accretion histories by applying a constant mass accretion rate from a given onset time. We varied both the onset times of mass accretion and the accretion rates to test the response of the convective core. Since these models describe accreting stars that can grow to roughly twice their initial mass during their MS evolution, the models effectively probe a stellar mass range of approximately 15--80 $M_\odot$. Figure \ref{fig:new_delta_prescription} shows the numerically derived values of $\delta$ from nine \textsc{Mesa} models of a 40 $M_\odot$ star undergoing mass gain under different conditions. Despite some scatter, the curves with different onset times and accretion rates all seem to show a universal trend. It is also clear that $\delta(M_c,Y_c)$ for mass accretors is not well aligned with the fitting function for $\delta(M_c,Y_c)$ derived in \citet{Shikauchi2024} for mass-losing stars.

We fit the models and propose an updated prescription of the function $\delta(Y_c)$ for MS accretors as
\begin{equation} \label{eq:new_delta_prescription}
    \delta(Y_c) = 2^{-\frac{Y_c - Y_0}{X_0}},
\end{equation}
where $X_0$ and $Y_0$ are the initial hydrogen and helium mass fractions. Our function strongly deviates from the \textsc{Mesa} results in the region $Y_c\gtrsim0.85$. The drop in the \textsc{Mesa} curves in this region is partially due to the uncertainties in the assumed $\alpha(M_c)$ when deriving the numerical $\delta(Y_c)$. However, small differences in the derivatives this late in the MS evolution will not make a significant difference on the final state of the MS star. This updated prescription for $\delta$ is valid in cases where accretion does not yet trigger stellar rejuvenation, which is treated separately and described in the next section.

\subsection{Rejuvenation of MS accretors} \label{sec:rejuvenation}
When an MS star accretes material from its companion at a sufficiently high mass accretion rate, the convective core grows and mixes in hydrogen-rich material. This lowers the central helium fraction, making the star appear younger and resulting in stellar rejuvenation. For rejuvenation to occur, the mass accretion rate needs to exceed the threshold given by 
\begin{equation}
    \dot{M}_\mathrm{threshold} = \frac{\alpha}{1-\alpha Y_c} \frac{M}{\beta \: \delta} \frac{L}{Q_\mathrm{CNO}M_c},
\end{equation}
reflecting that the mass gain must dominate the convective core mass change over the natural core shrinkage. To account for this situation, we propose a new stellar rejuvenation treatment for MS accretors. 

\begin{figure*}[t!]
    \plotone{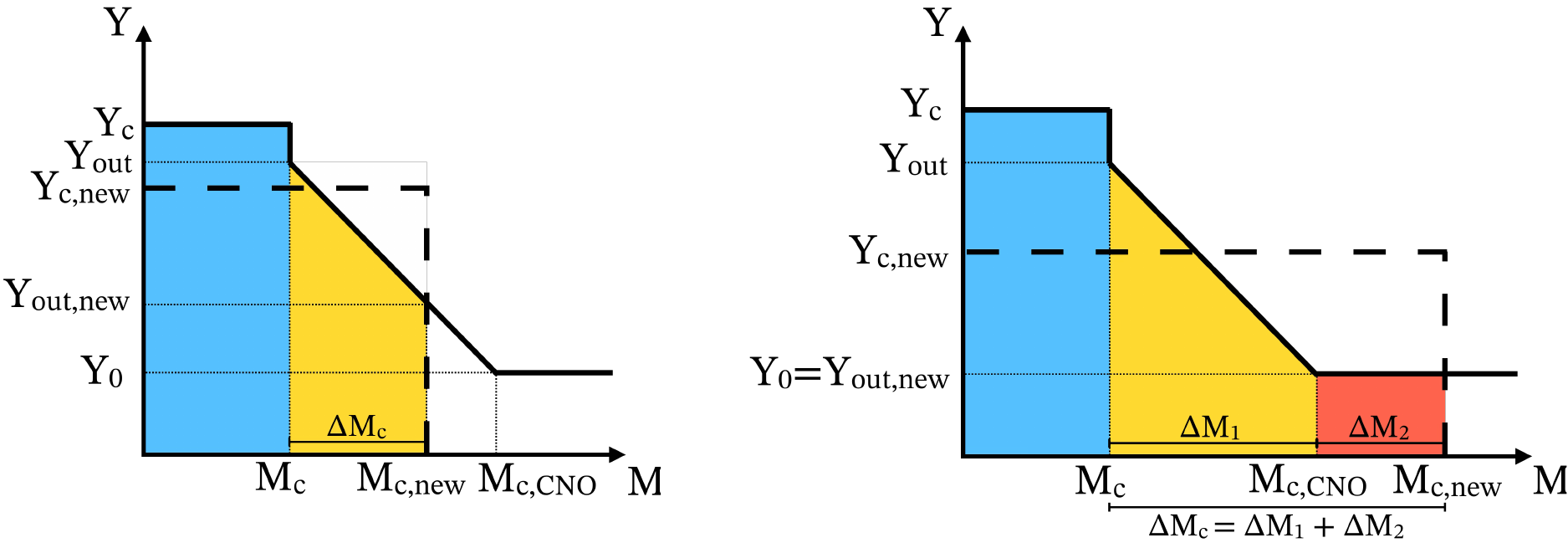}
    \caption{Graphical illustration of how the rejuvenation is treated in two cases: when the CNO-processed core mass $M_{c,\mathrm{CNO}}$ remains greater than the convective core mass after accretion (left), and when the convective core mass exceeds $M_{c,\mathrm{CNO}}$ (right). The solid black line represents the helium profile in the star before accretion, with the area underneath it corresponding to the helium mass. The dashed line represents the helium profile of the convective core after rejuvenation. The helium mass inside the expanded convective core (given as $Y_{c,\mathrm{new}}\times M_{c,\mathrm{new}}$) must equal the helium mass originally in the convective core plus the helium mixed in from the surrounding layers (the sum of the colored areas). The profiles and shaded areas are schematic and not drawn to scale.
\label{fig:core_not_exceeded}}
\end{figure*}

Consider an MS star with convective core mass $M_c$ and central helium fraction $Y_c$. During an episode of mass gain, the convective core mass of the accretor is predicted to increase by $\Delta M_c$. In our framework, this $\Delta M_c$ is obtained from Eq.~(\ref{eq:core_change}). However, if $\Delta M_c >0$, this procedure is inconsistent: using the $\Delta M_c$ obtained from Eq.~(\ref{eq:core_change}) together with the original $\Delta Y_c$ from Eq.~(\ref{eq:change_in_Yc_differential}) would neglect the rejuvenation, since fresh hydrogen is mixed into the core and $Y_c$ should decrease. We therefore need to recompute $\Delta Y_c$, and we make the assumption that the value of $\Delta M_c$ can be estimated from pre-accretion quantities using Eq.~(\ref{eq:core_change}). The total helium mass in the core after rejuvenation is
\begin{equation} \label{eq:new_helium_mass}
    M_{\rm{He, new}} = ( Y_c + \Delta Y_c) ( M_c + \Delta M_c),
\end{equation}
where $\Delta Y_c <0$ and $\Delta M_c > 0$. As the convective core grows, it expands into the region with previously processed material. To correctly estimate how much fresh hydrogen is mixed into the core, some information about the chemical profile outside the convective boundary is required. We assume a linear helium abundance profile between the convective core mass $M_c$ and the CNO-processed core mass $M_{c,\mathrm{CNO}}$, which represents the outermost mass coordinate within a star where hydrogen has been partially converted into helium through the CNO cycle. A linear distribution is expected for stars that evolve without mass loss, but the profile could deviate in cases with mass loss or gain. We use $Y_{\rm{out}}$ to denote the helium fraction of the gas just outside the convective core. Using the helium profile, we can calculate the change in the helium mass within the convective core as
\begin{equation}
    \Delta M_{\rm{He}} = Y_{\rm{out}} \Delta M_c - \frac{1}{2}\Delta M_c^2 \frac{Y_{\rm{out}} - Y_0}{M_{c,\mathrm{CNO}} - M_c},
\end{equation}
where $Y_{0}$ is the initial helium abundance. The total helium mass contained within the expanded convective core must equal the sum of the helium originally present in the core and the helium mixed in from the surrounding layers ($M_{\rm{He,new}} = Y_c M_c + \Delta M_{\rm{He}}$), so
\begin{equation} \label{eq:heliumconservation}
    M_{\rm{He,new}} = Y_c M_c + Y_{\rm{out}} \Delta M_c - \frac{1}{2}\Delta M_c^2 \frac{Y_{\rm{out}} - Y_0}{M_{c,\mathrm{CNO}} - M_c}.
\end{equation}
The helium profile inside the star before and after accretion is shown in Figure \ref{fig:core_not_exceeded}. Using Eq.~(\ref{eq:new_helium_mass}) and (\ref{eq:heliumconservation}) to solve for $\Delta Y_c$ gives
\begin{equation} \label{eq:deltaY_core_not_exceeded}
\Delta Y_c = \frac{\Delta M_c}{M_c + \Delta M_c} \left( Y_{\rm{out}} - Y_c + \frac{\Delta M_c (Y_{\rm{out}} - Y_0)}{2 \; (M_c - M_{c,\mathrm{CNO}})} \right).    
\end{equation}
We use linear interpolation to update the value of $Y_{\rm{out}}$ after rejuvenation
\begin{equation}  \label{eq:deltaYout_core_not_exceeded}
    Y_{\rm{out,new}} = Y_{\rm{out}} + \frac{Y_{\rm{out}} - Y_0}{M_c - M_{c,\mathrm{CNO}}} \cdot \Delta M_c.
\end{equation}
When the core does not grow ($\Delta M_c \leq 0$), $Y_\mathrm{out}$ is constantly updated as
\begin{equation}
    Y_{\rm{out}} = Y_c.
\end{equation}

Eq.~(\ref{eq:deltaY_core_not_exceeded}) is valid as long as the new convective core mass does not exceed the CNO-processed core mass ($M_{c,\mathrm{new}} \leq M_{c,\mathrm{CNO}}$). In the case that the convective core grows beyond the CNO-processed core, we use
\begin{equation} \label{eq:coreexceeded}
    M_{\rm{He, new}} =  Y_c M_c + \frac{1}{2} (Y_\mathrm{out} + Y_0) \Delta M_1 + \Delta M_2 Y_0,
\end{equation}
where $\Delta M_1 = M_{c,\mathrm{CNO}} - M_c$ is the mass that the convective core incorporates to reach the CNO-processed core boundary, and $\Delta M_2 = \Delta M_c - \Delta M_1$ is the additional material beyond that boundary, which will be mixed into the expanded convective core. Physically, $\Delta M_1$ corresponds to the material that is helium-enriched and was already partially processed, while $\Delta M_2$ corresponds to the previously unprocessed envelope material with the initial composition. The helium profiles before and after the convective core growth are illustrated in Figure \ref{fig:core_not_exceeded}. After obtaining $Y_{c,\mathrm{new}}=M_\mathrm{He,new}/M_{c,\mathrm{new}}$ using Eq.~(\ref{eq:coreexceeded}), we set $Y_{\rm{out,new}} = Y_0$ and $M_{c,\mathrm{CNO}} = M_c + \Delta M_c$, reflecting the fact that the CNO-processed core boundary has moved outward to the new convective core mass.

\section{Implementation} \label{sec:implementation}
While the \texttt{BRCEK} framework has been implemented on top of the \citet{Hurley2000} formalism for consistency within \textsc{Compas}, it effectively replaces the MS evolution for massive stars. The underlying framework is independent of the \citet{Hurley2000} stellar tracks and can be combined with other stellar models that provide post-MS evolution. In the \texttt{BRCEK} formalism, the convective core mass and luminosity evolve according to the framework of \citet{Shikauchi2024} and we use our rejuvenation model. For the stellar radius, we assume that it can be expressed as a function of the total stellar mass, central helium abundance, and surface helium abundance. Since we currently do not have analytical models describing the radius evolution of mass-losing and mass-gaining MS stars, we retain the functional form of the \citet{Hurley2000} radius prescription, but scale it based on the surface helium abundance to better capture the response of stars to mass loss.

This section addresses the challenges that arise when applying this new framework to the original \citet{Hurley2000} evolutionary tracks. We also outline additional features implemented as part of the \texttt{BRCEK} MS evolution formalism that improve the treatment of radius evolution, surface helium abundance, MS merger products, and spun-down chemically homogeneous stars.

\subsection{General core mass evolution} \label{sec:general_core_mass_evolution}
Evolution of the convective core's chemical composition and mass is governed by Eq.~(\ref{eq:change_in_Yc_differential}) and (\ref{eq:core_change}). We use the equations in an integrated form since \textsc{Compas} evolves stars using small timesteps. The timesteps are chosen to ensure that both components in binary stars change by no more than 0.1\% in mass due to winds and 10\% in radius due to stellar evolution. Timesteps can be further reduced if a star is close to filling its Roche lobe \citep{Mandel2025}. We assume that $\alpha(M_c)$ and $\delta(M_c,Y_c)$ can be approximated using their values at the beginning of the timestep, and that the changes in both the convective core mass and central helium fraction are small between the timesteps. This works even for larger changes in core mass if they happen on a thermal timescale as the core does not react on timescales this short. After taking a small timestep $\Delta t$, we calculate the change in central helium fraction by integrating Eq.~(\ref{eq:change_in_Yc_differential}) as
\begin{equation} \label{eq:changeinYc}
    \Delta Y_c = \frac{L}{Q_{\mathrm{CNO}}M_c}\;\Delta t.
\end{equation}
Similarly, we can integrate Eq.~(\ref{eq:core_change}) and obtain the change in convective core mass after the timestep as 
\begin{equation} \label{eq:changeinMc}
    \Delta M_c = \Delta M_{c,\mathrm{nat}} + \Delta M_{c,\mathrm{ML}},
\end{equation}
where $\Delta M_\mathrm{c,nat}$ is the natural decay (without mass loss) of the convective core over the timestep, given by
\begin{equation}
    \Delta M_{c,\mathrm{nat}} = -\frac{\alpha M_c}{1-\alpha Y_c}\;\Delta Y_c.
\end{equation}
The additional change in convective core mass due to mass loss is given by $\Delta M_\mathrm{c,ML}$. Here we need to take the change in $\beta(M)$ into account when integrating, since the total mass $M$ can change significantly over a single timestep (e.g. during mass transfer on a thermal timescale). The integrated equation can be written as
\begin{equation}
    \frac{\Delta M_{c,\mathrm{ML}}}{M_c} = \left( \frac{\Delta M}{M} + \frac{\Delta f}{f(M)} + \frac{\Delta M \;\Delta f}{M\;f(M)} \right) \; \delta(M_c,Y_c),
\end{equation}
where $\Delta M$ is the amount of mass lost, $Mf(M)$ represents the CNO-processed core mass at zero-age main sequence (ZAMS) for a star with mass $M$, $\Delta f = f(M + \Delta M) - f(M)$. The fitting function $f(M)$, describing the fraction of the convective core mass at ZAMS, is defined in \citet{Shikauchi2024}.

Our framework provides the MS evolution of massive stars, but for the post-MS phases we currently rely on the stellar tracks from \citet{Hurley2000}. To ensure a smooth and self-consistent transition between the two regimes, we must reconcile how the post-MS evolution is initialized, particularly the value of the helium core mass at TAMS. In the \citet{Hurley2000} formalism, this is controlled through parameter $M_0$. Although referred to as an “effective initial mass" by \citet{Hurley2000}, $M_0$ is used as a variable that sets the helium core mass once the envelope begins to decouple from the core. During the MS, $M_0$ is set equal to the total stellar mass $M$, but during later evolutionary stages $M_0$ is held fixed while $M$ continues to change with mass loss. For our purposes, $M_0$ must instead be set to reflect the convective core mass produced by our framework.

When the \texttt{BRCEK} framework is enabled, the convective core mass is calculated at each timestep during the MS. The helium core mass at the start of the Hertzsprung gap (HG) phase depends on the current $M_0$, so we update $M_0$ to a value consistent with the calculated convective core mass at TAMS using a bracketing root-finding method as implemented in the Boost C\texttt{++} library. We set the updated parameter $M_0$ as
\begin{equation}
    M_{0,\mathrm{new}} = \max (M_{0,\mathrm{solver}},\;\min (M, M_\mathrm{ZAMS})).
\end{equation}
This ensures physical consistency in both mass loss and mass gain scenarios. For stars that lose mass ($M < M_\mathrm{ZAMS}$), the condition prevents $M_0$ from dropping below the current total mass $M$, which would otherwise cause an unphysical luminosity drop during the stellar contraction phase at the MS hook. This issue is particularly relevant for intermediate-mass stars (5-10 $M_\odot$) due to the luminosity treatment in this mass range (see Section \ref{UpdatedStellarTracks}). For mass gain cases ($M > M_\mathrm{ZAMS}$), the default formalism tends to overestimate the helium core mass at TAMS. We therefore allow $M_0$ to drop below $M$, but never below $M_\mathrm{ZAMS}$, ensuring reasonable stellar evolution tracks for these stars.

Additionally, we impose a maximum convective core mass at 90\% of the total stellar mass. This choice is motivated by our \textsc{Mesa} models, which suggest that the convective core mass is never $>90\%$ of the total mass, regardless of how much mass is lost. This cap prevents MS stars from becoming marked as fully convective.

The \texttt{BRCEK} MS framework is applied to all stars with $M_\mathrm{ZAMS} \geq 1.5\;M_\odot$. This threshold is defined in terms of the ZAMS mass rather than the current stellar mass in order to avoid transitions between radiative and convective core structures during the MS, which are not explicitly modeled in the present framework. In the \citet{Hurley2000} formalism, the transition between stars with radiative and convective cores occurs at $M\approx0.95$--$1.12\,M_\odot$, depending on metallicity. Adopting a higher threshold of $1.5\,M_\odot$ therefore provides a conservative lower limit that ensures stars treated within the \texttt{BRCEK} framework retain convective cores even if they experience moderate mass loss during the MS. For stars with $M_\mathrm{ZAMS} < 1.5\:M_\odot$, the detailed convective core mass and rejuvenation calculations are disabled, and the \texttt{MANDEL} method is used instead. While it is not a detailed model, this fallback still preserves higher helium core masses at TAMS following case A mass transfer.

\subsection{Rejuvenation}
We implement the rejuvenation routine as follows. At each timestep:
\begin{enumerate}
    \item Compute changes: We calculate the change in central helium fraction $\Delta Y_c$ and convective core mass $\Delta M_c$ using Eq.~(\ref{eq:changeinYc}) and (\ref{eq:changeinMc}), following \citet{Shikauchi2024}.
    
    \item No core growth ($\Delta M_c \leq 0$): If the core does not grow, we update the values directly $M_{c,\mathrm{new}} = M_c + \Delta M_c$, $Y_{c,\mathrm{new}} = Y_c + \Delta Y_c$, and $Y_{\mathrm{out,new}} = Y_{c,\mathrm{new}}$.
    
    \item Core growth ($\Delta M_c > 0$): If the core grows, we apply rejuvenation by adjusting $\Delta Y_c$. Two cases are considered:  
    
    Convective core mass does not exceed the CNO-processed core mass ($M_c + \Delta M_c \leq M_{c,\mathrm{CNO}}$): We use Eq.~(\ref{eq:deltaY_core_not_exceeded}) to obtain $\Delta Y_c$ and calculate the new central helium fraction $Y_{c,\mathrm{new}} = Y_c + \Delta Y_c$.  Eq.~(\ref{eq:deltaYout_core_not_exceeded}) is used to determine the new helium fraction just outside the core $Y_{\rm{out,new}}$.

    Convective core mass exceeds the CNO-processed core mass ($M_c + \Delta M_c > M_{c,\mathrm{CNO}}$): We set $Y_{c,\mathrm{new}}=M_{\mathrm{He,new}}/M_\mathrm{c,new}$, where $M_{\mathrm{He,new}}$ is computed via Eq.~(\ref{eq:coreexceeded}). We then set the helium fraction just outside the core $Y_{\rm{out,new}} = Y_0$, and update the CNO-processed core mass $M_{c,\mathrm{CNO}} = M_{c,\mathrm{new}}$.
\end{enumerate}

\subsection{Surface helium abundance}
Detailed tracking of the convective core mass and helium profile inside the star also enables more detailed modeling of the surface helium abundance in stripped MS stars. An MS star begins its life with an initial helium abundance $Y_0$ and initial convective core mass $M_c$, which is initially the same as $M_{c,\mathrm{CNO}}$, corresponding to the initial CNO-processed core boundary. As the star evolves, the convective core gradually recedes, leaving behind layers of partially processed material with elevated helium abundance $Y>Y_0$. If the star undergoes significant mass loss, such that the total mass drops below the CNO-processed core mass, the helium-enriched material will be exposed at the surface. In this case, the surface helium abundance $Y_s$ will no longer reflect the initial value $Y_0$ and will be enhanced. Assuming a linear helium profile in mass across the region between the receded convective core and the CNO-processed core boundary, we can determine the updated surface helium abundance as
\begin{equation} \label{eq:surface_helium_abundance}
    Y_{s,\mathrm{new}} = Y_\mathrm{out} + (M_\mathrm{new} - M_c) \frac{Y_s - Y_\mathrm{out}}{M_{c,\mathrm{CNO}} - M_c},
\end{equation}
where $M_\mathrm{new}$ is the total mass after stripping and $M_\mathrm{new} \leq M_{c,\mathrm{CNO}}$. This linear assumption is generally a good approximation for stars with no mass loss. However, we note that there will be deviations when the star has complex mass-change histories. The only time this deviation becomes relevant is when a star that has had a complex mass-loss history later on starts accreting mass at rates high enough to cause rejuvenation. We expect such situations to be rather rare in regular binary evolution.

The implementation proceeds as follows: As long as the total stellar mass after the timestep satisfies $M_\mathrm{new} > M_{c,\mathrm{CNO}}$, the surface helium abundance $Y_s$ is set to the initial abundance $Y_0$. If the total mass drops below $M_{c,\mathrm{CNO}}$ after taking the timestep, the surface helium abundance is updated using Eq.~(\ref{eq:surface_helium_abundance}). The CNO-processed core mass is then reset to the new total mass, $M_{c,\mathrm{CNO}}=M_\mathrm{new}$, to reflect the fact that the original CNO-processed core boundary has changed. 

In addition to following the evolution of the surface helium abundance, our framework also tracks the surface hydrogen abundance, given by $X_s=1-Y_s-Z$. While the framework captures the evolution of $X_s$ and $Y_s$ for mass-losing MS stars, the treatment of mass-gaining MS stars currently assumes that the accreted material has the same composition as the star's initial surface abundance. Consequently, accretion of helium-enriched material from a stripped MS star or evolved companion is not reflected in the accretor's surface abundances.

\subsection{Luminosity} \label{UpdatedStellarTracks}
\citet{Shikauchi2024} provide a luminosity prescription as a function of the convective core mass $M_c$ and central helium fraction $Y_c$. This prescription is applied whenever the \texttt{BRCEK} MS evolution framework is enabled and the ZAMS mass satisfies $M_\mathrm{ZAMS} \geq 15 \;  M_\odot$. It is valid only during the core-hydrogen burning phase (i.e. $Y_c < 1 - Z$) and does not capture the MS hook (evolution during the MS hook is described in Section \ref{sec:MS_hook}). 

For intermediate-mass stars with $M_\mathrm{ZAMS} < 15\;M_\odot$, we adopt the original luminosity prescription from \citet{Hurley2000}. It is important to note that the \citet{Hurley2000} luminosity prescription depends on the total stellar mass and age, while the \citet{Shikauchi2024} prescription is based on the convective core mass and central helium fraction. As a result, these different dependencies can lead to significantly different luminosity evolution after a mass transfer episode.

\subsection{Radius}
We currently do not have a prescription that describes the radius of stars that have experienced mass changes. This remains a complex problem, as the radius is highly sensitive to the star's internal chemical profile \citep{Shikauchi2024}. For this reason, we keep the original radius prescription from \citet{Hurley2000}, but introduce several important modifications to improve physical consistency.

In the stellar tracks from \citet{Hurley2000}, radius is a function of the effective age $t_\mathrm{eff}$, total mass, and metallicity, $R_\mathrm{Hurley}=R_\mathrm{Hurley}(t_\mathrm{eff},M;Z)$. In the framework of \citet{Shikauchi2024}, the central helium fraction $Y_c$ serves as a proxy for stellar age, gradually increasing throughout the core-hydrogen-burning phase. To be able to use the original MS radius prescription, we need to convert the central helium fraction to an effective age. At each timestep, we linearly scale the effective age with the central helium fraction as
\begin{equation}
    t_\mathrm{eff} = 0.99 \; t_{\mathrm{MS}} \; \frac{Y_c - Y_0}{X_0},
\end{equation}
where $X_0$ and $Y_0$ are the initial hydrogen and helium abundances, and $t_{\mathrm{MS}}=t_\mathrm{MS}(M;Z)$ is the MS lifetime from \citet{Hurley2000}. This adjustment is applied only during the first 99\% of the star's MS evolution, reflecting the phase of core hydrogen burning. The factor of 0.99 ensures that this correction does not extend into the final 1\% if the MS lifetime, during which the star has already exhausted its central hydrogen and undergoes a brief stellar contraction phase (see Section \ref{sec:MS_hook}).

An additional modification involves scaling the stellar radius based on the surface helium abundance. As long as the total mass remains above the CNO-processed core mass $M_{c,\mathrm{CNO}}$, the surface helium abundance $Y_s$ stays at the initial value $Y_0$, and we use the standard \citet{Hurley2000} radius formula. However, once the star loses enough mass that $M<M_{c,\mathrm{CNO}}$, we update $Y_s$  using Eq.~(\ref{eq:surface_helium_abundance}), and apply a modified radius prescription
\begin{equation} \label{eq:modified_radius}
    R = R_\mathrm{Hurley} + (R_\mathrm{ZAMS} - R_\mathrm{Hurley}) \cdot \frac{Y_s - Y_0}{Y_c - Y_0},
\end{equation}
where $R_\mathrm{Hurley}$ is the standard radius from \citet{Hurley2000}, which depends only on the star's metallicity, mass, age, and is unaffected by the new luminosity prescription or convective core mass. This scaling reflects the fact that, once the outer hydrogen-rich envelope is removed or substantially reduced, partially stripped stars remain much more compact \citep[e.g.][]{Laplace2020}. The interpolation ensures that the stellar radius transitions smoothly from the default \citet{Hurley2000} value toward the ZAMS radius as surface helium increases due to stripping. It is also consistent with the implementation of chemically homogeneous stars in \textsc{Compas}, which are fully mixed ($Y_s = Y_c$) and their radius is set to the ZAMS radius \citep{Riley2021}.

\begin{figure*}[ht!]
\plotone{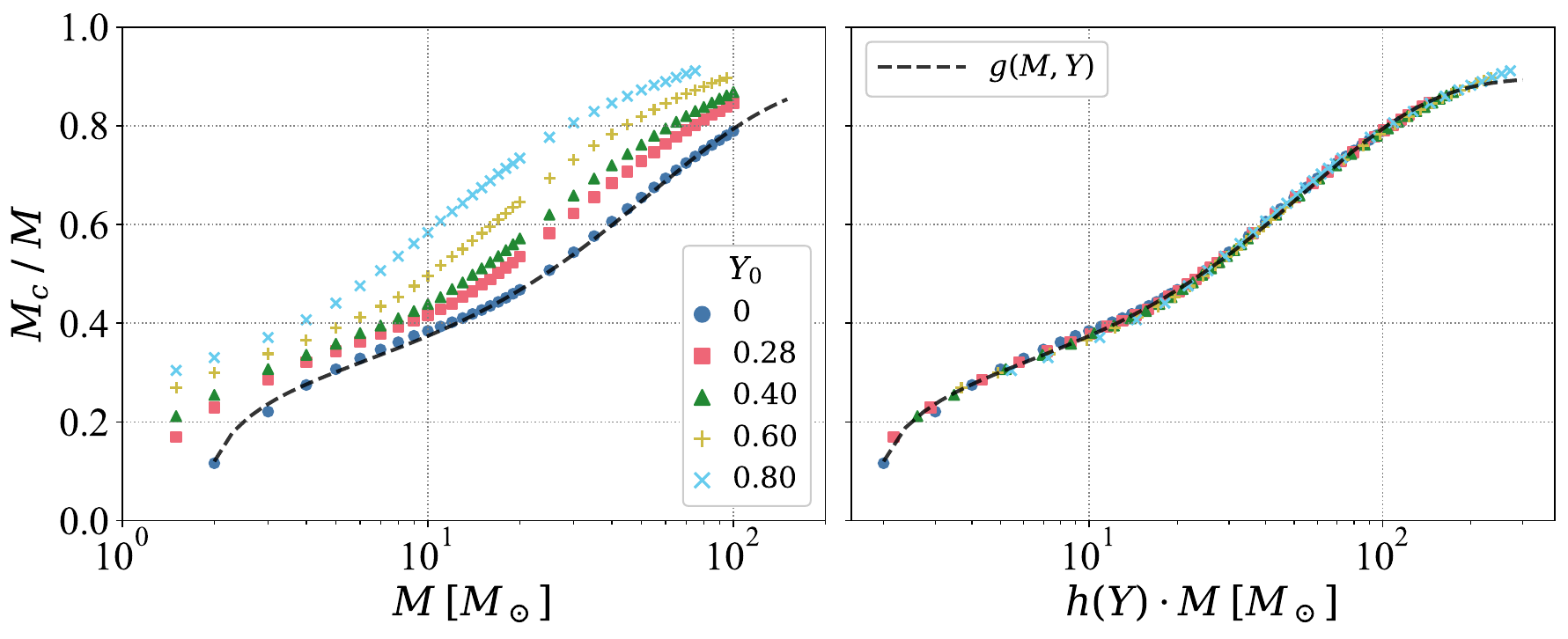}
\caption{Initial core-to-total mass ratio as a function of total stellar mass from \textsc{Mesa} models with varying initial helium abundances. When the total mass is scaled by $h(Y)$ (see Eq.~\ref{eq:h(Y)}), all points lie on the same curve, described by the fitting function $g(M,Y)$ (see Eq.~\ref{eq:fmix_fitting_function}). This function predicts the initial convective core mass (initial CNO-processed core mass) based on the star's total mass and initial helium abundance, and is used to determine the convective core mass of uniformly mixed merger products and spun-down chemically homogeneous stars.
\label{fig:new_fmix}}
\end{figure*}

\subsection{Evolution during the MS hook} \label{sec:MS_hook}
Stars with convective cores experience a brief phase of stellar contraction near the end of the MS phase, during which their radius decreases, and effective temperature and luminosity increase before transitioning to the HG. In a Hertzsprung-Russell (HR) diagram (luminosity vs. effective temperature space), this manifests as a ``hook" feature at the end of MS evolution. To account for this phase, we use linear interpolation to connect the luminosity at the end of core hydrogen burning to the start of the HG phase
\begin{equation} \label{eq:luminosity_interpolation}
    L(t) = \frac{L_\mathrm{hook} \cdot (t_\mathrm{MS} - t) + L_\mathrm{HG,0} \cdot (t - t_\mathrm{hook})} {t_\mathrm{MS} - t_\mathrm{hook}},
\end{equation}
for $t_\mathrm{MS} \geq t \geq t_\mathrm{hook}$, and where $L_\mathrm{hook}$ is the luminosity at the onset of the hook, $L_\mathrm{HG,0}$ is the luminosity at the beginning of the HG (with the updated $M_0$ value, see Section \ref{sec:general_core_mass_evolution}), and $t_\mathrm{hook}=0.99\:t_\mathrm{MS}$ \citep{Hurley2000} is the time at which the hook phase begins.

Similar to the treatment of luminosity during the MS hook phase, we use linear interpolation to ensure a smooth transition in radius between the exhaustion of core hydrogen and the start of the HG evolution. This interpolation follows a formula analogous to Eq.~(\ref{eq:luminosity_interpolation}), with the TAMS radius defined as $R_\mathrm{TAMS} = \min (R, R_\mathrm{HG,0})$, where $R$ is the current radius and $R_\mathrm{HG,0}$ is the radius at the beginning of the HG phase. This choice prevents unphysical expansion during the MS hook that could otherwise trigger premature mass transfer. If the star has been significantly stripped during the MS phase, its radius may fall below $R_\mathrm{HG,0}$. In such cases, we keep the current radius fixed during the hook, deferring any potential mass transfer episode until the HG, when the star has developed a clear core-envelope structure in the \citet{Hurley2000} formalism and mass transfer can be more accurately handled.

\subsection{MS mergers and chemically homogeneous evolution}
MS stars can undergo additional evolutionary pathways. They may merge, or if they are massive and rapidly rotating, they may undergo chemically homogeneous evolution (CHE). In the latter case, rotationally induced mixing transports material from the convective core throughout the radiative envelope, maintaining a uniform composition in the star \citep{Maeder1987}. The treatment for CHE in \textsc{Compas} is described in \citet{Riley2021}. 

To improve the treatment of MS mergers and CHE outcomes, we use a set of \textsc{Mesa} models to estimate the initial convective core mass as a function of the initial helium abundance. We find that the initial convective core mass for a given stellar mass $M$ and initial helium mass fraction $Y$ can be described via the following fitting function
\begin{equation} \label{eq:fmix_fitting_function}
    g(M, Y) = \left( c_1 + c_2 \cdot e^{-\frac{Mh(Y)}{ c_3 M_\odot}} \right) \cdot \left(1 - \frac{c_5 M_\odot}{M h(Y)} \right)^{c_4},
\end{equation}
where
\begin{equation} \label{eq:h(Y)}
    h(Y) = 10^{\frac{Y(Y+2)}{4}},
\end{equation}
with the fitting coefficients being
\begin{eqnarray}
    c_1 &=& 0.89817, \; c_2 = -0.59224, \; c_3 = 55.78853, \nonumber \\
    c_4 &=& 0.35908, \; c_5 = 1.87718. \nonumber
\end{eqnarray}
The initial convective core mass, corresponding to the initial CNO-processed core mass, is then defined as $M_{c,\mathrm{CNO}} = M \cdot g(M,Y)$. Figure \ref{fig:new_fmix} shows the results from \textsc{Mesa} models and the fitting function $g(M,Y)$.

The fitting function is applicable to both chemically homogeneous stars that eventually spin down and evolve like normal MS stars, and to MS mergers, assuming that the merger product is uniformly mixed. In \textsc{Compas}, MS mergers are assumed to lose a fraction of mass during coalescence, based on the results of \citet{Wang2022}. By assuming the lost mass originates from the hydrogen-rich envelope, and by using the helium profile of the progenitor stars, we can estimate the total hydrogen mass retained in the merger product. This allows for a more accurate modeling of the composition and subsequent evolution of the newly formed, homogeneously mixed star.

\section{Results} \label{sec:results}

\subsection{Single star evolution}

\begin{figure*}[t!]
\plotone{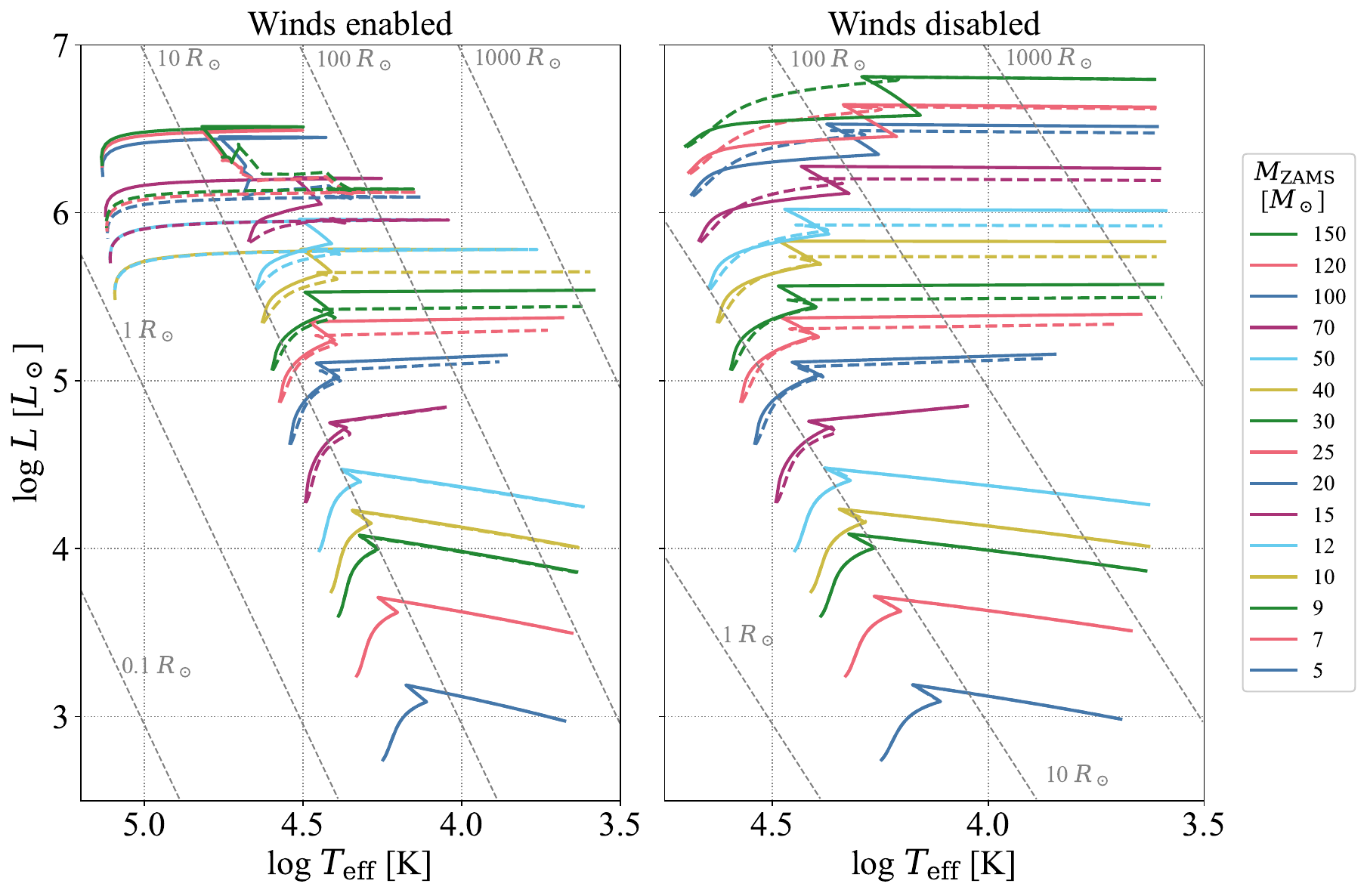}
\caption{Hertzsprung-Russell diagram showing stellar tracks for various initial masses at $Z=Z_\odot$ with mass loss via stellar winds from \citet{Merritt2025} enabled (left) and mass loss disabled (right). Only the MS and HG evolution is shown. Solid lines show the updated stellar tracks when \texttt{BRCEK} formalism is used, and dashed lines are the default stellar tracks from \citet{Hurley2000}. The luminosity prescription from \citet{Shikauchi2024} is enabled only for stars with $M_\mathrm{ZAMS} \geq 15\;M_\odot$.
\label{fig:HRdiagram}}
\end{figure*}

Since massive MS stars lose a lot of mass via stellar winds, our new framework has an impact on single star evolution. We compare the new stellar evolution tracks with the original tracks from \citet{Hurley2000} in Figure \ref{fig:HRdiagram}. The luminosity prescription from \citet{Shikauchi2024} (as a part of the \texttt{BRCEK} framework) results in systematically higher luminosities during the MS. The most significant differences arise for high-mass stars experiencing strong wind-driven mass loss. In these cases, the \texttt{BRCEK} MS framework results in substantially larger helium core masses at TAMS, increasing the post-MS luminosity. This higher luminosity can further enhance the wind mass-loss rates, as seen in the $40\:M_\odot$ model that self-strips on the HG and becomes a naked helium star in our framework, whereas the same star does not self-strip under the original \texttt{HURLEY} formalism. Even in the absence of wind mass loss, the \texttt{BRCEK} framework modifies both the MS luminosity of massive stars and their TAMS helium core mass, leading to differences in post-MS luminosity that are visible in the high-mass tracks. These differences may partly reflect variations in the adopted mixing prescriptions and the extent of convective overshooting (see Section \ref{sec:CC_mass_evolution}).

\begin{figure*}[ht!]
\plotone{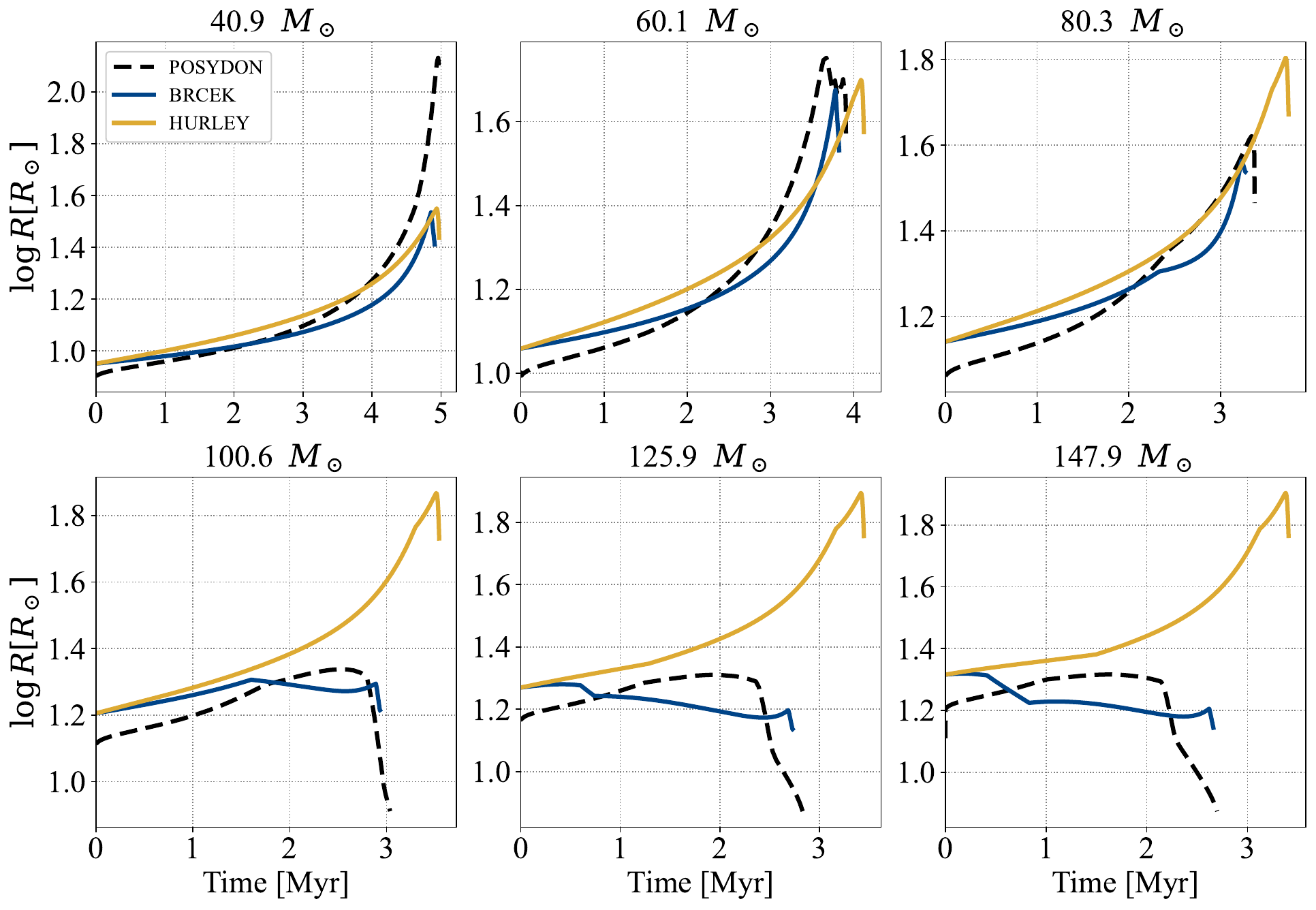}
\caption{Radial evolution of MS stars ($Z=Z_\odot$) losing mass via stellar winds, shown for the \textsc{Mesa}-based \textsc{Posydon} models (dashed black lines), the \texttt{BRCEK} MS evolution framework with modified radius (solid blue), and the original radius prescription from \citet{Hurley2000} (solid yellow). The ZAMS mass is indicated above each panel. The impact of our modified radius prescription (Eq.~\ref{eq:modified_radius}) becomes visible in the 80.3 $M_\odot$ model, where the radial evolution changes around 2.5 Myr. 
\label{fig:RadialEvolutionComparison}}
\end{figure*}

Figure \ref{fig:RadialEvolutionComparison} shows the radial evolution of MS stars for three approaches: the original \texttt{HURLEY} prescription, our \texttt{BRCEK} framework with modified radius using Eq.~(\ref{eq:modified_radius}), and the \textsc{Mesa}-based single-star hydrogen-burning models \citep{andrews_2025_15194708} used in the \textsc{Posydon} binary population synthesis framework \citep{Fragos2023,Xing2023,Andrews2024A}. This comparison is intended to be purely qualitative, as the models rely on different physical assumptions, including the adopted wind mass-loss prescriptions and the treatment of convective boundary mixing. The original \citet{Hurley2000} treatment of MS evolution tends to overestimate the radius for stripped massive stars compared to detailed models. The \texttt{BRCEK} framework with the modified radius prescription addresses this problem and yields MS lifetimes that are also closer to those of \textsc{Posydon} models. In the 40.9 $M_\odot$ model, the \textsc{Posydon} stellar tracks yield larger radii due to the different assumptions in the models. The difference between the \texttt{BRCEK} and \texttt{HURLEY} frameworks arises primarily from the different luminosity prescription (see Section \ref{UpdatedStellarTracks}), which affects the wind-driven mass loss and hence the stellar radius. The scaling in Eq. (\ref{eq:modified_radius}) only reduces the stellar radius relative to the \texttt{HURLEY} prescription and does not provide a mechanism to increase it.

\begin{figure*}[ht]
\plotone{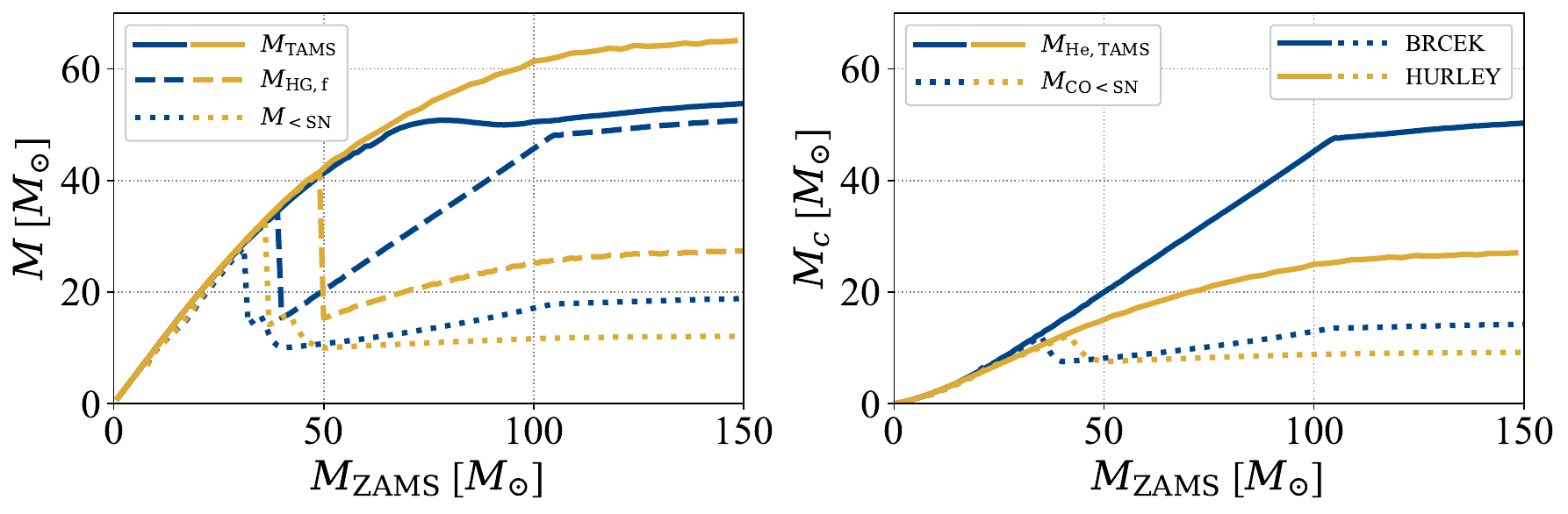}
\caption{The left panel shows the total stellar mass at TAMS ($M_\mathrm{TAMS}$, solid lines), end of HG evolution ($M_\mathrm{HG,f}$, dashed), and just before the supernova ($M_\mathrm{<SN}$, dotted) as a function of the ZAMS mass ($Z=Z_\odot$). The right panel shows the helium core mass at TAMS ($M_\mathrm{He,TAMS}$, solid lines) and carbon-oxygen core mass just before the supernova ($M_\mathrm{CO<SN}$, dotted). The \texttt{BRCEK} framework is shown in blue and the \texttt{HURLEY} prescription in yellow. We use the default prescription for stellar winds from \citet{Merritt2025}.
\label{fig:CoreMassvsInitialMassSSE}}
\end{figure*}

 Figure \ref{fig:CoreMassvsInitialMassSSE} shows the total mass of single stars as a function of the initial mass at three evolutionary stages: at TAMS, at the end of the HG phase, and just before the core collapse. We also show the helium core mass at TAMS and carbon-oxygen core mass just before the supernova. The \texttt{BRCEK} framework results in significantly larger helium core masses at TAMS compared to the \texttt{HURLEY} prescription. However, much of this additional helium core mass is lost during post-MS evolution due to strong stellar winds, particularly in the most massive stars. The larger helium core mass also leads to a dramatically different core-to-envelope mass ratio at the start of the HG. For stars with $M_\mathrm{ZAMS} > 100\;M_\odot$, the \texttt{HURLEY} formalism yields an envelope mass comprising approximately 55\% of the total mass, whereas in \texttt{BRCEK} models, the envelope accounts for only 10\% (set by our limit for the maximum convective core mass, see Section \ref{sec:general_core_mass_evolution}). The strong mass-loss rate between TAMS and the end of the HG phase (visible for $M_\mathrm{ZAMS} \gtrsim 40\;M_\odot$) is caused by the onset of the luminous blue variable phase \citep{Humphreys1994}, characterized by very strong winds that remove the star's hydrogen envelope.

\begin{figure*}[ht]
    \plotone{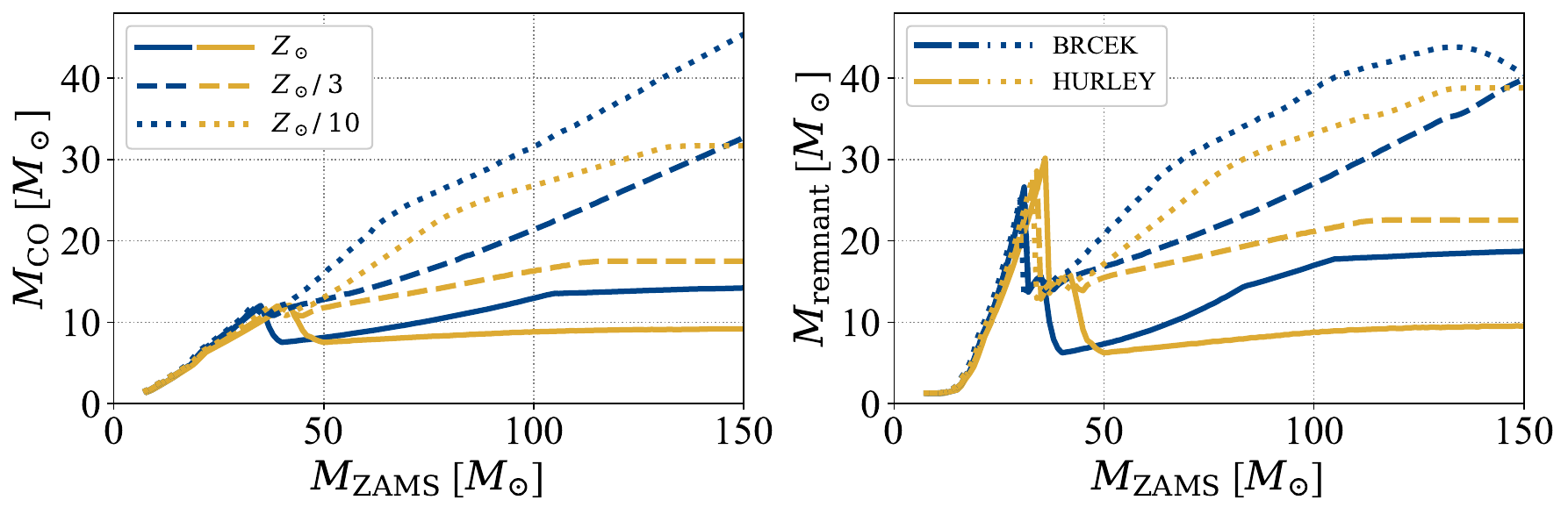}
    \caption{The left panel shows the carbon-oxygen core mass just before core collapse as a function of the ZAMS mass for three different metallicities, comparing the \texttt{BRCEK} (blue) and \texttt{HURLEY} (yellow) approaches. The right panel shows the corresponding remnant masses, calculated using the remnant mass prescription from \citet{Fryer2012} and pulsational pair-instability supernova prescription from \citet{Marchant2019}. All models use the default stellar wind prescription from \citet{Merritt2025}.
    }
    \label{fig:RemMass_vs_InitMass}
\end{figure*}

Our new framework has direct consequences on the mass of the final compact object. Figure \ref{fig:RemMass_vs_InitMass} shows the carbon-oxygen core mass prior to core collapse and the corresponding remnant mass as a function of the star's initial mass, calculated using the stellar wind prescription from \citet{Merritt2025} and the remnant mass prescription from \citet{Fryer2012} to avoid stochasticity in the final remnant masses. The results are shown for three metallicities: $Z_\odot$, $Z_\odot/3$, $Z_\odot / 10$. The \texttt{BRCEK} framework leads to higher remnant masses for stars with $M_\mathrm{ZAMS} > 50\;M_\odot$ across all tested metallicities. The decline in remnant masses for the most massive stars at $Z=Z_\odot/10$ is due to the onset of pulsational pair-instability, modeled following the prescription of \citet{Marchant2019}.

\subsection{Binary star evolution}
Implications of our new framework on stellar populations will be explored in a future study (Brček et al. in prep.), but here we explore some direct consequences on binary star evolution. To evaluate the effects of the new MS evolution framework on case A mass transfer, we create a set of \textsc{Compas} binary star models with predefined initial configurations. \textsc{Compas} does not allow setting mass-transfer rate directly, so to ensure reproducibility of our results, we control mass transfer outcomes by varying initial separations of the stars.

\begin{figure}[t]
\plotone{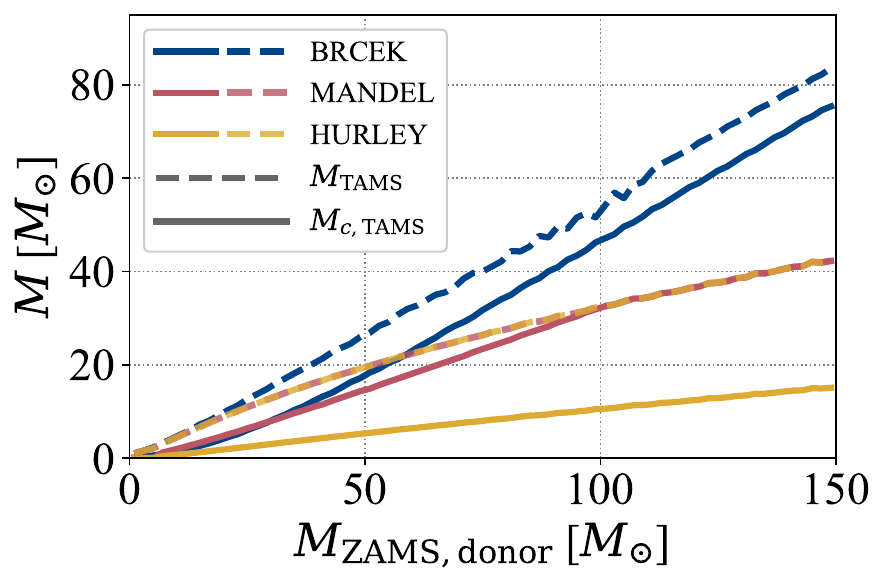}
\caption{Helium core mass of the donor star at TAMS after experiencing case A mass transfer (solid lines) and the total mass of the donor at TAMS (dashed lines) for three different MS evolution treatments. Initial separation was varied such that the primary star loses a significant fraction of its mass due to case A mass transfer. Stellar winds were disabled. The total stellar mass of the donor at TAMS is the same between \texttt{MANDEL} and \texttt{HURLEY} prescriptions.
\label{fig:CoreMassvsInitialMass}}
\end{figure}

For a given primary star of initial mass $M_{1,i}$, we set the initial secondary mass to $M_{2,i}=0.7M_{1,i}$, corresponding to the mass ratio of $q=0.7$. We choose the initial orbital separation as 
\begin{equation}
    a_i = 5\left(\frac{M_{1,i}}{M_\odot}\right)^{0.6} R_\odot,
\end{equation}
which ensures that the primary star fills its Roche lobe and initiates stable mass transfer in the second half of its MS lifetime. While the exact onset time and the amount of mass transferred vary with $M_{1,i}$, in all cases, the primary star loses a significant portion of its mass during the MS phase. In these sets of models, the mass loss via stellar winds is disabled and we assume solar metallicity ($Z_\odot=0.0142$). Figure \ref{fig:CoreMassvsInitialMass} shows the donor's helium core mass and total stellar mass at TAMS, as a function of the donor's initial mass for the three MS evolution treatments now available in \textsc{Compas}. The total mass at TAMS differs between the \texttt{BRCEK} and \texttt{HURLEY/MANDEL} methods, despite identical initial conditions. In the \texttt{MANDEL} and \texttt{HURLEY} prescriptions, the MS evolution is decoupled from the convective core mass, resulting in behavior that is insensitive to internal structure. While the \texttt{MANDEL} and \texttt{HURLEY} prescriptions predict the same total mass at TAMS, the \texttt{MANDEL} prescription predicts a larger helium core mass for stripped stars, up to the value of the total mass at TAMS. In contrast, the \texttt{BRCEK} framework directly links the convective core mass to luminosity and radius, altering the MS evolution of some systems. In our framework, once the star is stripped down to its CNO-processed core boundary, the surface helium abundance increases, leading to a suppression of the stellar radius according to Eq.~(\ref{eq:modified_radius}). This reduction in radius allows the donor to disengage from mass transfer earlier and retain more mass.

\begin{figure*}[ht!]
    \plotone{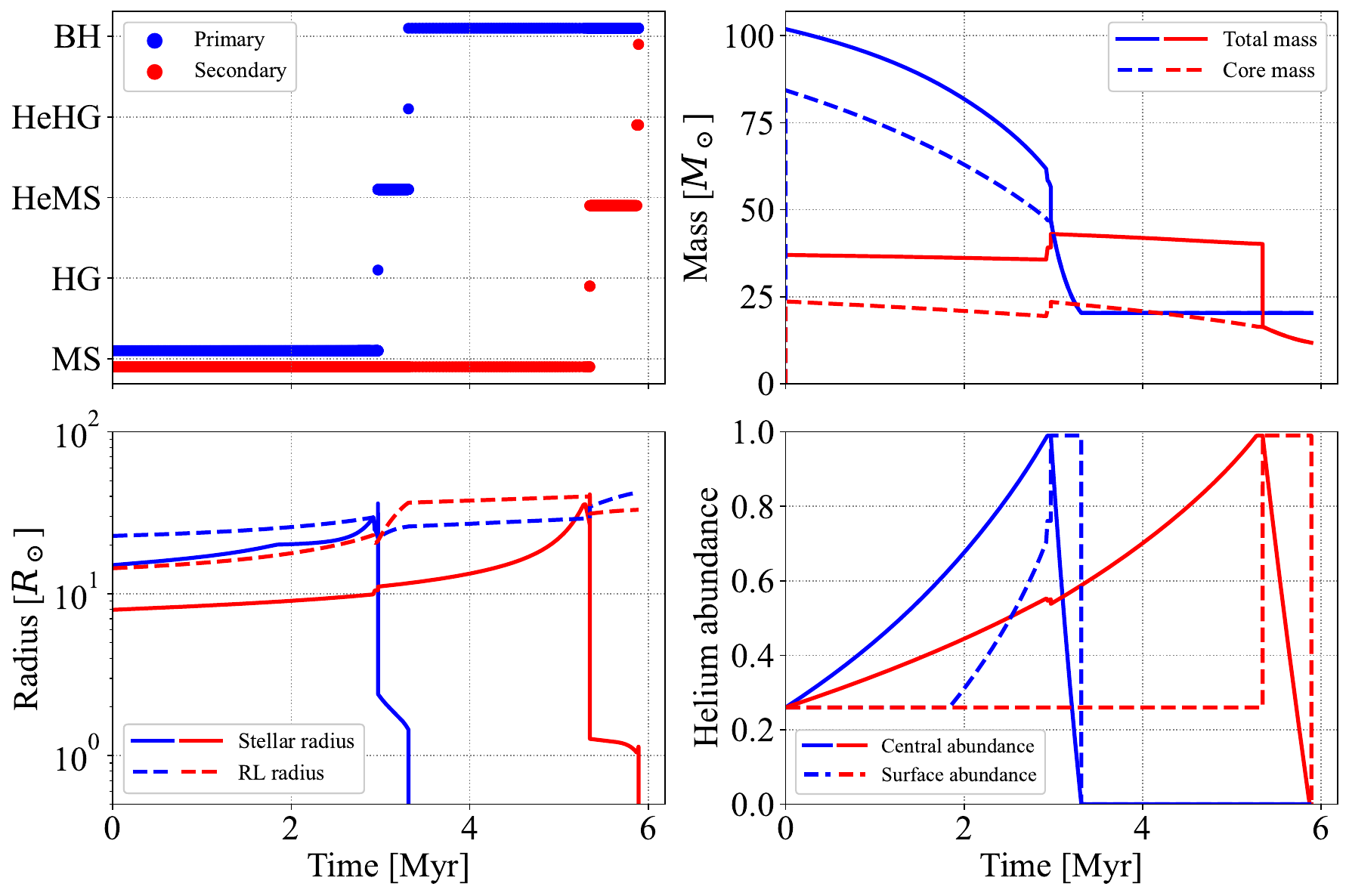}
    \caption{Detailed evolution of a representative binary system illustrating the effects of the \texttt{BRCEK} framework. The initial primary mass is $M_\mathrm{ZAMS,1}=101.9 \; M_\odot$, secondary mass $M_\mathrm{ZAMS,2}=37.0 \; M_\odot$, the initial separation is $a_i = 45.2 \; R_\odot$, and the metallicity is $Z=0.01$. Other options are kept at \textsc{Compas} default. Top left panel shows the stellar type as stars transition through MS, HG, helium main sequence (HeMS), helium Hertzsprung gap (HeHG), and black hole formation (BH). Other panels show the evolution of total mass and convective core mass (top right), radius and Roche lobe radius (bottom left), and central and surface helium fraction (bottom right). This system demonstrates how mass loss and accretion affect the radius, convective core mass, and chemical composition.
    }
    \label{fig:detailed_evolution}
\end{figure*}

To demonstrate the full framework in action, we show the detailed time evolution of a binary system in Figure \ref{fig:detailed_evolution}. The initial primary mass is 101.9 $M_\odot$, initial secondary mass is 37.0 $M_\odot$, initial separation is 45.2 $R_\odot$\ and the metallicity is sub-solar at $Z=0.01$. Just before 2~Myr, the primary's total mass decreases to the level of its CNO-processed core mass. As a result, partially processed layers become exposed at the surface, leading to an increase in surface helium abundance. At the same time, the radius expansion slows down, as controlled by our radius scaling prescription. Our framework allows for the tracking of surface hydrogen and helium abundances during the MS, enabling the identification of partially stripped MS stars with enhanced helium abundance (WNh stars, e.g. \citealt{Crowther2007}), which \textsc{Compas} was unable to do before. At around 3 Myr, the primary fills its Roche lobe, triggering stable mass transfer to the secondary. As the secondary accretes material, its convective core mass increases and its central helium fraction decreases, indicating partial rejuvenation. This rejuvenation effect is less pronounced compared to the \citet{Hurley2000} framework, particularly when accretion occurs later in the MS, when the core is less sensitive to changes in the total mass. Note that the surface helium abundance of the accreting secondary does not track the addition of helium-enriched material in our simplified model.

This system is particularly interesting, because at $\sim$5~Myr into its evolution, it produces a system similar to Cygnus X-1 -- a high mass X-ray binary (HMXB), consisting of a $~21.2 \; M_\odot$ BH and a $~40.6\;M_\odot$ MS star that is almost Roche lobe filling \citep{Miller-Jones2021}. One aspect of the configuration that is difficult to reproduce is the orbital period: Cygnus X-1 system has an orbital period of $\sim$5.6 days, while our reproduced system has a longer period of $\sim$10 days.

To evaluate the population-scale impact of our framework on the MS evolution of massive stars, we generate a synthetic population of binary stars at solar metallicity using the default \textsc{Compas} configuration. Primary masses are drawn from the initial mass function of \citet{Kroupa2001}, mass ratios from a uniform distribution, and orbital separations from a flat distribution in logarithmic space. To focus only on the effects of binary evolution, we disable stellar winds in all models. We also allow the evolution of MS merger products (in default \textsc{Compas} settings, evolution of a binary stops after an MS merger). We produce a series of HR diagram snapshots over time and compile them into an animation to visualize the MS evolution of the stellar population. Figure \ref{fig:animation} shows a representative frame from this animation, comparing the results under the \texttt{HURLEY} and \texttt{BRCEK} MS evolution treatments. 

One clear feature is the emergence of blue straggler stars (BSSs), stars that extend beyond the MS turn-off \citep{Sandage1953}. BSSs formed through MS mergers and mass transfer in the original \citet{Hurley2000} framework tend to be uniformly distributed in temperature in the BSS region of the HR diagram. In contrast, our framework produces a more distinct separation between merger products and accretors. The most luminous BSSs form almost exclusively through MS mergers and occupy the hotter region of the HR diagram. The difference arises from the \citet{Shikauchi2024} luminosity prescription, which depends on the star's convective core mass and central helium fraction. Although the accretors' convective core masses grow during mass transfer, the increase is not sufficient to significantly boost their luminosity in most cases.

\section{Discussion}
The HR diagrams in Figure \ref{fig:HRdiagram} illustrate the differences in stellar tracks for single stars under different MS evolution frameworks. For stars with $M_\mathrm{ZAMS}>100 \; M_\odot$, stellar tracks appear irregular, but it is important to note that the radius prescription still relies on the underlying \citet{Hurley2000} formalism, which itself is based on fits to stellar models only up to 50 $M_\odot$ \citep{Pols1998}. For higher initial masses, the tracks represent extrapolations that may not accurately capture the evolution. This issue is reflected in the predicted stellar radii, shown in Figure \ref{fig:RadialEvolutionComparison}. The \citet{Hurley2000} models significantly overestimate the maximum radius of stripped massive stars compared to detailed \textsc{Posydon} models. Our framework mitigates this issue by scaling the radius based on surface helium abundance once the CNO-processed core is exposed. Developing a radius prescription that is fully consistent with the internal evolution in the \texttt{BRCEK} framework is ongoing work. The radius evolution is particularly relevant for modeling the mass transfer in close binaries, where it critically determines when Roche-lobe overflow occurs.

The stellar models underlying the \citet{Shikauchi2024}, \citet{Pols1998}, and POSYDON grids adopt different treatments of convective boundary mixing. This then affects stellar parameters and differences between the models are expected. The strength of our framework lies in parametrization, which is not tied to a specific \textsc{Mesa} model. The framework defines a functional description of the convective core evolution under mass change, while the numerical coefficients encode the calibration to a given set of stellar models. In principle, the same parametrization could be recalibrated using stellar tracks with different assumptions. We leave the exploration of sensitivity to different mixing prescriptions for future studies.

\begin{figure}[t!]
\begin{interactive}{animation}{HRD_evolution.mp4}
\vspace{0.5cm}
\plotone{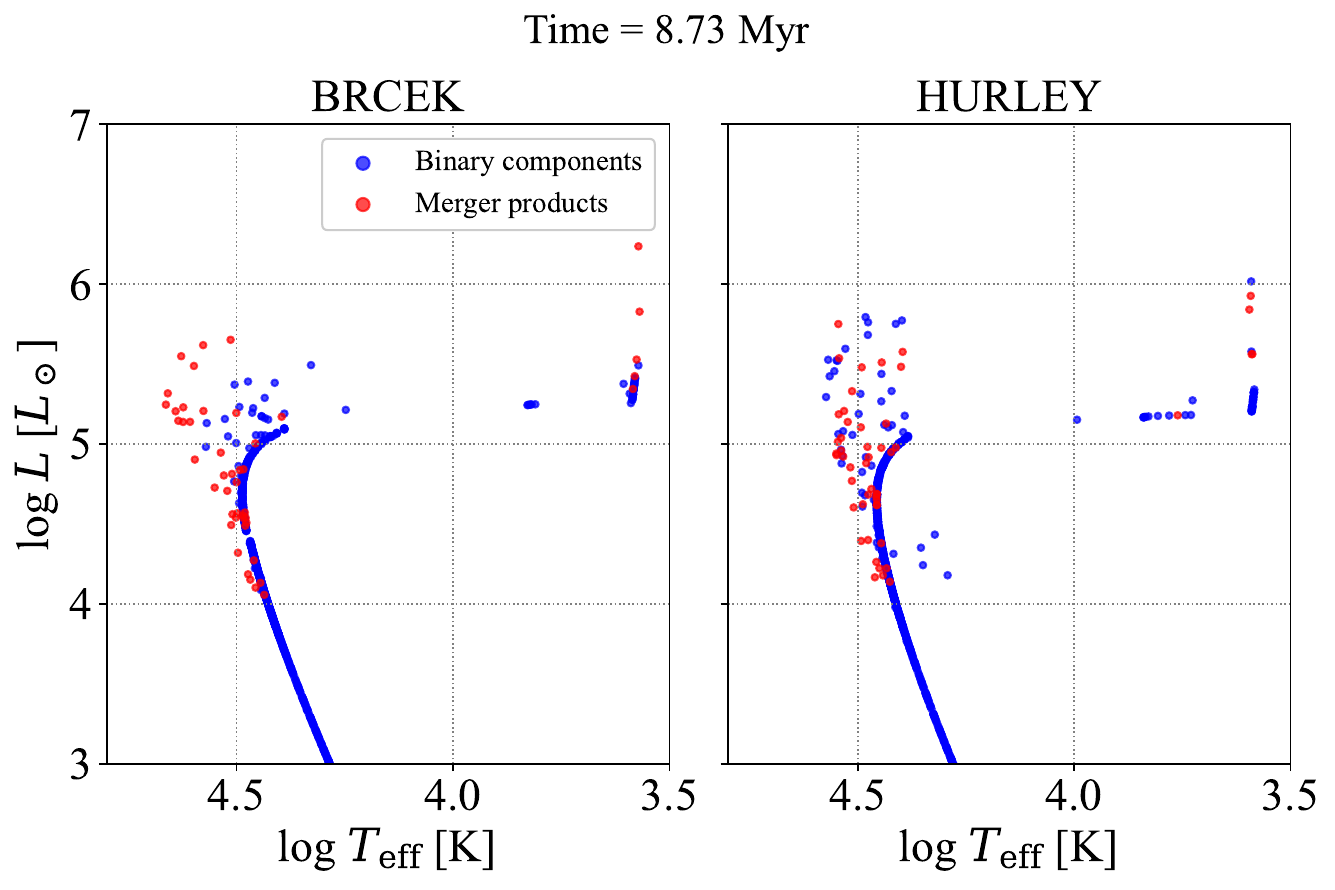} 
\caption{Animation showing the time evolution of a binary population in the HR diagram over the first 30 Myr of evolution after ZAMS. The snapshot is taken at 8.73 Myr. The population was generated using the default \textsc{Compas} configuration at solar metallicity, and stellar winds were disabled. The two panels compare our \texttt{BRCEK} MS evolution framework and the original \texttt{HURLEY} formalism. \label{fig:animation}
}
\end{interactive}
\end{figure}

The luminosity evolution differs significantly between the \texttt{BRCEK} and \texttt{HURLEY} treatments, not only for single stars (Figure \ref{fig:HRdiagram}) but also for post-mass-transfer binaries. This is evident in the animation shown in Figure \ref{fig:animation}, where BSSs formed through mass transfer and mergers occupy distinct regions of the HR diagram. Additionally, a cluster of stars appears to the right of the MS (in the blue giant region). In the \texttt{HURLEY} panel, this cluster is present from early times, whereas in our framework it emerges only towards the end of the animation. This is a consequence of the luminosity prescription. We apply our convective core mass calculations to stars below 15 $M_\odot$ but retain the original \citet{Hurley2000} luminosity prescription. If a star undergoes mass transfer near the end of its MS lifetime, its core mass remains largely unaffected, and its luminosity should also remain stable. However, because the \citet{Hurley2000} luminosity depends on the total mass rather than the convective core mass, the luminosity drops as the star loses mass, which then artificially extends its MS lifetime and produces this clustering. A unified luminosity prescription applicable across the full mass range would eliminate this artifact.

An additional limitation arises near the lower-mass boundary of our framework, where the evolution of convective cores can become non-monotonic and is sensitive to the adopted treatment of convective boundary mixing \citep[e.g.][]{Deheuvels2016}. The present framework is calibrated primarily on more massive stellar models and its application below $\sim 10 M_\odot$ should be regarded as an extrapolation.

A key consequence of our new MS evolution framework is the higher helium core mass and significantly altered core-to-envelope mass ratio at TAMS, particularly in mass-losing stars as shown in Figure \ref{fig:CoreMassvsInitialMassSSE}. These more massive cores also lead to systematically larger carbon-oxygen core masses prior to core collapse, as well as higher final remnant masses, as shown in Figure \ref{fig:RemMass_vs_InitMass}. The altered remnant mass distribution is expected to impact the chirp mass distribution that gravitational-wave detectors observe. This will be explored in future work (Brček et al. in prep.).

The system shown in Figure \ref{fig:detailed_evolution} is particularly interesting due to its HMXB phase around 5 Myr, when the ~40 $M_\odot$ MS secondary becomes nearly Roche lobe filling, while the primary has already evolved into a 20 $M_\odot$ BH. Such a configuration cannot be produced under the original \texttt{HURLEY} framework, highlighting the importance of our new treatment of the convective core evolution. The inability to reproduce the observed orbital period of Cygnus X-1 may point to the need to include tidal interactions or to an overestimation of stellar radii in our models. The formation and properties of HMXBs under this new framework will also be the subject of future work (Brček et al. in prep.).

\section{Conclusion}
We have presented a new MS evolution framework for massive stars, implemented in the rapid binary population synthesis code \textsc{Compas}. Our framework captures the evolution of the convective core on the MS, building on the semi-analytical model of \citet{Shikauchi2024}. We extend the original model to also account for mass gain through a physically motivated treatment of stellar rejuvenation. We derive a fitting function for the initial convective core mass as a function of the helium abundance, enabling more accurate modeling of MS merger products and spun-down CHE stars. Our treatment also results in more compact radii in mass-losing MS stars, more closely resembling detailed models, and enables modeling of surface helium and hydrogen abundances for MS stars stripped down to their CNO-processed core mass. 

When applied to binary systems, our model alters both the mass and composition structure of donors and accretors. The rejuvenation treatment incorporates the internal structure of stars, resulting in more realistic MS lifetimes and core growth for accretors. Compared to the original \citet{Hurley2000, Hurley2002} formalism, our framework yields higher helium and carbon-oxygen core masses, leading to more massive compact remnants. These changes are expected to impact predictions for the BH mass spectrum and the properties of gravitational-wave sources. Our model also enables the formation of HMXB systems that resemble Cygnus X-1 under the default \textsc{Compas} mass transfer prescriptions. An exploration of the consequences of this new framework, particularly on BH-HMXB and binary BH populations, will be presented in a future study (Brček et al. in prep.). 

Overall, this updated treatment of the MS evolution introduces a physically motivated connection between stellar structure and predicted stellar properties, improving the realism of rapid binary population synthesis predictions.

\begin{acknowledgments}
We thank Jeff Riley, Simon Stevenson, Dorottya Szécsi, and Jan Eldridge for helpful discussions. We also thank the anonymous referee for their comments that have improved the manuscript. We acknowledge support from the Australian Research Council (ARC) Centre of Excellence for Gravitational Wave Discovery (OzGrav), through project number CE230100016. A.B. acknowledges support from the Commonwealth through an Australian Government Research Training Program Scholarship [DOI: \url{https://doi.org/10.82133/C42F-K220}]. R.H. acknowledges support from the RIKEN Special Postdoctoral Researcher Program for junior scientists.
\end{acknowledgments}

\bibliography{bibliography}{}
\bibliographystyle{aasjournalv7}



\end{document}